\documentclass[a4paper,fleqn,usenatbib]{mnras}

\usepackage[T1]{fontenc}
\usepackage{ae,aecompl}

\usepackage{graphicx}	
\usepackage{amsmath}	
\usepackage{amssymb}	

\title[RFM for FRBs]{Three aspects of the radius-to-frequency mapping in fast radio bursts}

\author[H. Tong et al.]{H. Tong$^{1}$\thanks{E-mail: tonghao@gzhu.edu.cn},
J. Liu$^{1}$, H. G. Wang$^{1}$, 
Z. Yan$^{2}$
\\
$^{1}$School of Physics and Materials Science, Guangzhou University, Guangzhou 510006, China\\
$^{2}$Shanghai Astronomical Observatory, Chinese Academy of Sciences, 80 Nandan Road, Shanghai 200030, China
}
\date{Accepted XXX. Received YYY; in original form ZZZ}

\pubyear{2015}

\begin{document}
\label{firstpage}
\pagerange{\pageref{firstpage}--\pageref{lastpage}}
\maketitle

\begin{abstract}
We further explored the radius-to-frequency mapping in cases of FRBs. An analytical treatment of Lyutikov (2020) is presented. The frequency dependence of the drifting rate and the drifting timescale are obtained. The aberration effect and the twist of the magnetic field lines may result in drifting in both directions. For one FRB, the burst width is larger at lower frequency according to the radius-to-frequency mapping. For the FRB population, the magnetic fields of the repeaters may be larger than that of the non-repeaters. Then, according to the radius-to-frequency mapping, the burst widths of the repeaters will be wider than that of the apparent non-repeaters. If similar window function (or emission cones) like that of pulsars and magnetars is also at work in the case of FRBs, then the window function may explain the single or multiple components of FRB profiles. The radius-to-frequency mapping modeling is to some degree independent of the underlying radio emission mechanism.
\end{abstract}



\section{Introduction}

\begin{figure}
\centering
\includegraphics[width=0.45\textwidth]{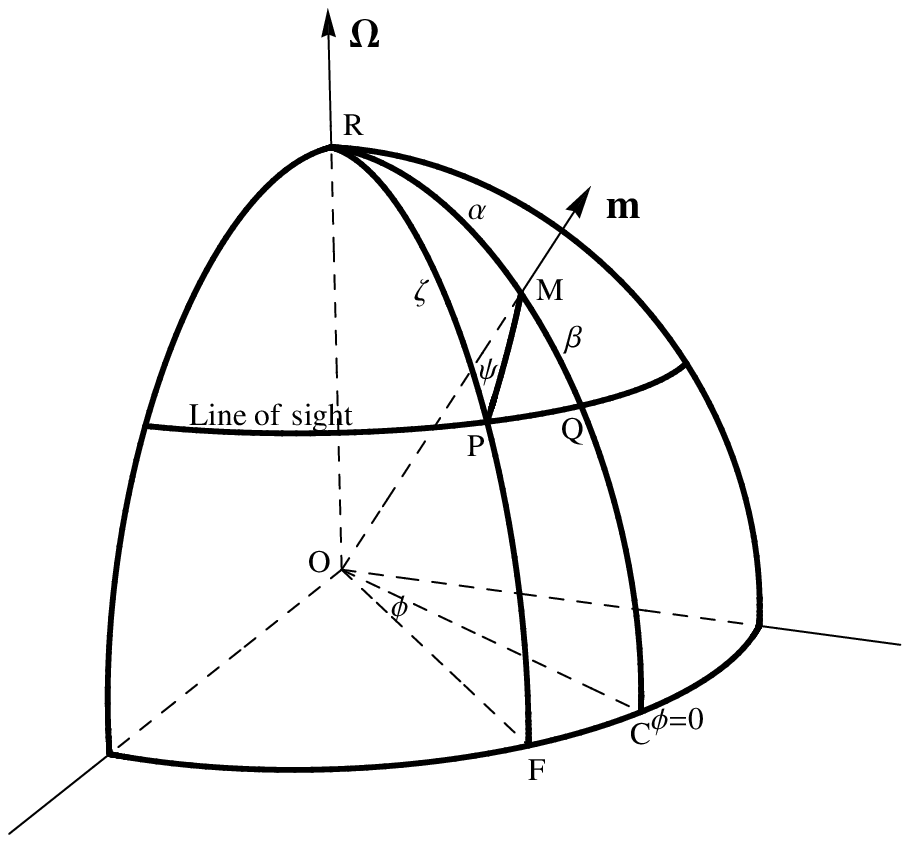}
\caption{Geometry of a rotating magnetosphere of a pulsar or a FRB. The rotational axis, the magnetic axis and the line of sight are shown. The angle between the rotational axis and the magnetic axis is the inclination angle $\alpha$. The angle between the rotational axis and the line of sight is denoted as $\zeta$. The closest approach between the line of sight and the magnetic axis is the impact angle $\beta$. From the figure, it can be seen that $\beta = \zeta-\alpha$. Note that $\beta$ can be either positive or negative. The $\psi$ is the position angle of the magnetic axis relative to the rotational axis, as seen from point $P$. The $\phi$ is the rotational phase. The arc $\overline{PM}$ is the colatitude of the observational point in the magnetic frame $\theta_{\rm obs}$. (Adapted from figure 1 in Tong et al. 2021)}
\label{fig_gmagnetosphere}
\end{figure}

Fast radio bursts (FRBs) are short duration radio flashes (Lorimer et al. 2007; Thortron et al. 2013). Their origins are not clear at present (Petroff et al. 2019; Zhang 2020). Due to their short durations ($\sim \rm ms$) and high brightness temperature, they are often thought to be coherent radio emissions from compact objects (neutron stars or black holes). Magnetars may account for some of the FRBs (Popov \& Postnov 2007; Katz 2016; Beloborodov 2017; Margalit \& Metzger 2018). For example, the Galactic magnetar SGR 1935+2154 showed radio bursts like FRBs (CHIME/FRB collaboration 2020; Bochenek et al. 2020; Lin et al. 2020; Kirsten et al. 2021). A possible sub-second periodicity was also found in one FRB (CHIME/FRB collaboration 2021b).
Both of the polarization position angle swing (Luo et al. 2020) and the sub-second periodicity (CHIME/FRB collaboration 2021b) support a magnetospheric origin of FRBs . A rotating magnetosphere like that of a pulsar is assumed for FRBs in the following context, see figure \ref{fig_gmagnetosphere}. This is the basic assumption of this paper.

There are different approaches to FRBs, like the synchrotron maser from the magnetar blast waves and so on (Lyubarsky 2014; Beloborodov 2017; Metzger et al. 2019; Beloborodov 2020; see Lyubarsky 2021 for a review). However, as based on the assumption of a rotating magnetosphere, the magnetospheric models are mainly considered here (Kumar et al. 2017; Katz 2018; Yang \& Zhang 2018).

Coherent radio emissions of pulsars have been studied for more than 50 years (Lyne \& Graham-Smith 2012). The uncertainties of pulsar radio emissions also hinder our understanding on FRBs (Lyubarsky 2021). At the same time, some geometrical models of pulsars have been established: the rotating vector model (Radhakrishnan \& Cooke 1969; Komeroff 1970), the radius-to-frequency mapping (RFM, Ruderman \& Sutherland 1975; Cordes 1978; Phillips 1992; Gangadhara \& Gupta 2001; Dyks et al. 2004), the beam shape (cone+core, Rankin 1983; patchy beam, Lyne \& Manchester 1988; fan beam, Wang et al. 2014) etc. These geometrical models are to some degree independent of the underlying radio emission mechanism. And they can explain many aspects of pulsar pulse profiles. This motivated us to apply the geometrical models of pulsars to FRBs. We focus on the RFM here.

Previously, the RFM was applied to explain the frequency-down drifting of repeaters (observations in Hessels et al. 2019; CHIME/FRB collaboration 2019a,b; Fonseca et al. 2020; theoretical modeling in Wang et al. 2019; Lyutikov 2020). We found that the following three aspects may all be related to the effect of the RFM: (1) Some repeaters showed down-drifting in the time-frequency domain, (2) The burst width is larger at lower frequency (Michilli et al. 2018; Pleunis et al. 2021a), (3) The burst widths of the repeaters tend to be larger than that of the apparent non-repeaters (CHIME/FRB collaboration 2019b; Fonseca et al. 2020; CHIME/FRB collaboration 2021a).

In Section 2, the frequency dependence of the drifting rate and the drifting timescale are obtained, through an analytical treatment of previous works. Secction 3 discussed possible effects that may result in drifting in both directions, including the aberration effect and the twist of the magnetic field lines. Section 4 discussed the possible consequences of the RFM for one FRB and the FRB population etc.  The conclusion is given in Section 5. In each item, the previous experiences of pulsars are firstly presented, then the discussions of FRBs are made.

\section{Frequency dependence of drifting rate and drifting timescale}

\subsection{Previous works}

In the case of pulsars, a steady magnetosphere may be assumed for the integrated pulse profile. In the scheme of the RFM, different frequencies come from different altitudes: $\nu \propto r^{-\alpha}$ (Ruderman \& Sutherland 1975; Cordes 1978; Gil et al. 1984; Phillips 1992; Wang et al. 2013). To be more specific, the lower frequencies are emitted at the higher altitudes. Then, it is generally expected that: (1) the pulse is wider at the lower frequency, (2) low frequency pulses will arrive earlier (they lie nearer to the observer, for a steady magnetosphere). Both of point (1) (Chen \& Wang 2014) and point (2) (Gangadhara \& Gupta 2001; Dyks et al. 2004) have been seen in the case of pulsars.

The point (1) may remain in cases of the FRBs (see discussions in section \ref{section_RFM}). However, FRBs may have a sudden spark instead of a steady magnetosphere.  Therefore, low frequency emissions will arrive later because they are emitted later (Wang et al.2019).

In Wang et al. (2019), the authors considered two separations between the high frequency emission region and the low frequency emission region: (1) radial separation $\Delta r$, (2) angular separation $\Delta \phi$ (both in the magnetic frame). Straight forward consideration shows that, the time delay due to the radial separation is:
\begin{eqnarray}
  \Delta t_r &=& \frac{\Delta r}{v} - \frac{\Delta r}{c} \approx \frac{\Delta r}{2\gamma^2 c} \\
  &\approx& 2\times 10^{-8} {\ \rm s} \left( \frac{\Delta r}{10^8 \ \rm cm} \right) \left( \frac{300}{\gamma} \right)^2,
\end{eqnarray}
where $v$ and $\gamma$ are the velocity and the Lorentz factor of the electrons, respectively.
This timescale is much smaller than the timescale of the frequency down-drifting in FRB ($\sim \rm ms$).
Geometrical consideration shows that the time delay due to the angular separation is\footnote{The angular width occupies a segment of $\sin\theta_{\rm obs} \Delta \phi$ on a small circle with circumference $2\pi \sin\zeta$, see figure \ref{fig_gmagnetosphere}. The neutron star sweeps the small circle in a rotational period $P$. A correction is made compared with Wang et al. (2019), eq.(7) there. The factor should be $\sin\theta_{\rm obs}$ instead of $\sin\beta$. Only when the line of sight lies in the plane of the rotational axis and the magnetic axis does $\theta_{\rm obs} = \zeta-\alpha=\beta$. The calculation is similar to the aberration effect in the case of pulsars, see discussion in section \ref{section_aberration}.}:
\begin{equation}\label{eqn_dt_phi}
  \Delta t_{\phi} = \frac{\sin\theta_{\rm obs} \Delta \phi}{2\pi \sin\zeta} P,
\end{equation}
where $P$ is the rotational period of the neutron star, $\theta_{\rm obs}$ is the colatitude of the observational point in the magnetic frame, $\zeta$ is the angle between the rotational axis and the line of sight, see figure \ref{fig_gmagnetosphere}. In Wang et al. (2019), the angular separation $\Delta \phi$ is a free parameter and the down-drifting timescale is mainly determined by this parameter.

Later work by Lyutikov (2020) made two improvements: (1) The author pointed correctly that the down-drifting of FRBs should be related to the RFM of pulsars (instead of the drifting subpulse discussed in Wang et al. 2019), (2) The curvatures of the magnetic field lines should be considered. Then, the radial separation will account for the time delay instead of an angular free parameter\footnote{This angular free parameter may still be at work in reality. But its exact value and physical mechanism is unknown at present.}. Following the procedure in Lyutikov (2020), the emission altitude, the observer time, and the frequency can all be calculated as functions of the time. It is consistent with the general expectation of Wang et al. (2019): the emissions at lower frequencies arrive later because they are emitted later. However, the procedure is rather complicated. In the case of small angle approximation, explicit analytical expression can be obtained. This will make clear the physics involved.

\subsection{Small angle approximations: the frequency dependence of the drifting rate and the drifting timescale}

An analytical treatment of Lyutikov (2020) is given in the following in the case of small angle approximations.
The field line equation of a dipole magnetic field is: $r = r_{e} \sin^2\theta$, where $r_e$ is the maximum radial extent of the magnetic field line. The distance along a field line from $\theta_0$ to $\theta$ is:
\begin{equation}
  \Delta s = \int \sqrt{(dr)^2 + r^2 (d\theta)^2} = \frac{r_0}{\sin^2\theta_0} \int_{\theta_0}^{\theta} \sqrt{1+ 3\cos^2\theta} \sin\theta d \theta,
\end{equation}
where $r=r_0$ when $\theta=\theta_0$ (spherical polar angle in the magnetic frame when $t=0$). At the time $t$, the distance that the relativistic particle travelled (particles can only slide along the field lines) is: $\Delta s (t) = c t$, where the velocity of the particle is assumed to be $v \approx c$. During one FRB, a flux tube may be ignited. If the angular scale of the flux tube is small, then differences between different field lines in the flux tube can be neglected. The geometry of all field lines can be represented by the field line with $r_0$ and $\theta_0$. For the case of small $\theta$, the integration in $\Delta s$ can be performed analytically. At the time $t$, the particle moves to the angle $\theta(t)$:
\begin{equation}
  \Delta s(t) \approx r_0 \left[ \left( \frac{\theta(t)}{\theta_0} \right)^2 -1 \right].
\end{equation}
Therefore, the emission altitude at the time $t$ is:
\begin{equation}
  r(t) = r_0 \frac{\sin^2\theta(t)}{\sin^2 \theta_0} \approx r_0 + \Delta s(t) = r_0 + c t.
\end{equation}
What's more, the observer time $t_{\rm obs}$ is a function of the coordinate time $t$ (Lyutikov 2020):
\begin{equation}\label{eqn_tobs}
  t_{\rm obs} = t- \frac{r(t) - r_0}{c} \cos(\theta_{\rm obs} - \theta_{\rm em}) \approx t[1 - \cos(\theta_{\rm obs} -\theta_{\rm em})],
\end{equation}
where $\theta_{\rm em}$ is the colatitude of the emission point in the magnetic frame. This equation refers to a burst extended in magnetic colatitude and $t_{\rm obs}$ refers to different electrons following different magnetic field lines (see figure 1 in Lyutikov 2020 for details). The change of the observer time may be written as:
\begin{equation}
  \Delta t_{r, \rm curvature} = \frac{r(t) - r_0}{c} \cos(\theta_{\rm obs} - \theta_{\rm em}).
\end{equation}
This expression (for radial separation considering curvature effect) is similar to other factors, e.g., in
eq.(\ref{eqn_dt_phi})(due to angular separation), eq.(\ref{eqn_dt_aberration})(due to the aberration effect), and eq.(\ref{eqn_dt_twist})(due to the twist of the magnetic field lines). These factors will be treated item by item.
For small $\theta_{\rm obs}$, it can be obtained that: $\theta_{\rm em} =(2/3) \theta_{\rm obs}$ (see Appendix C in Tong et al. 2021). This relation will be used later.
The emission altitude as a function of the observer time is:
\begin{eqnarray}
\label{eqn_r_tobs}
  r(t_{\rm obs}) &=& r_0 + c t = r_0 + c t_{\rm obs} \frac{1}{1-\cos(\theta_{\rm obs} - \theta_{\rm em})} \\
  &=& r_0 + c t_{\rm obs} f(\theta_{\rm obs}),
\end{eqnarray}
where the angular factor is: $f(\theta_{\rm obs}) = 1/(1-\cos(\theta_{\rm obs} - \theta_{\rm em}))$.

A general modeling of the RFM can be summarized in a power law form:
\begin{equation}
\label{eqn_RFM}
  \nu \propto r^{-\alpha},
\end{equation}
which states that the emission frequency is a function of the emission altitude. For curvature radiation, $\alpha=1/2$ (Wang et al. 2013). For plasma frequency, $\alpha=3/2$ (Ruderman \& Sutherland 1975). Then, the emission frequency as a function of the observer time is:
\begin{eqnarray}
  \frac{\nu(t_{\rm obs})}{\nu_0} &=& \left( \frac{r_0 + c t_{\rm obs} f(\theta_{\rm obs})}{r_0} \right)^{-\alpha}\\
  &=& \left( 1+ f(\theta_{\rm obs}) \frac{c t_{\rm obs}}{r_0} \right)^{-\alpha}
\end{eqnarray}
where $\nu_0$ is the frequency of the emission from the height of $r_0$ at the beginning of the burst ($t=0$). Using the previous simplifications, it can can be clearly seen now that the frequency will be lower at later time, i.e., the down-drifting in the time-frequency domains (Hessels et al. 2019; Josephy et al. 2019; CHIME/FRB collaboration 2019a, b; Fonseca et al. 2020; Pleunis et al. 2021b).

The drifting rate can be obtained by the time derivative of the frequency:
\begin{equation}\label{eqn_nudot}
  \frac{\dot{\nu}}{\nu_0} = -\alpha \left( \frac{\nu}{\nu_0} \right)^{1+1/\alpha} f(\theta_{\rm obs}) \frac{c}{r_0}.
\end{equation}
If the observational band is rather narrow ($\nu \approx \nu_0$), then (note that $\alpha$ is approximately 1):
\begin{equation}
  \frac{\dot{\nu}}{\nu} \approx -\alpha f(\theta_{\rm obs}) \frac{c}{r_0} \approx - f(\theta_{\rm obs}) \frac{c}{r_0}.
\end{equation}
The typical drifting time scale will be:
\begin{eqnarray}
 \frac{\nu}{|\dot{\nu}|} &\approx& \frac{1}{f(\theta_{\rm obs})} \frac{r_0}{c} \\
  &=& [1- \cos(\theta_{\rm obs} - \theta_{\rm em})] \frac{r_0}{c}.
\end{eqnarray}
Using the relation between $\theta_{\rm obs}$ and $\theta_{\rm em}$, for an observational colatitude $\theta_{\rm obs} \leq 1 (\rm rad)$,
the typical drifting timescale is:
\begin{equation}
 \tau_{\rm drifting} = \frac{\nu}{|\dot{\nu}|} \sim \frac{1}{18} \frac{r_0}{c} = 1 \ {\rm ms} \left(\frac{r_0}{5\times 10^8 \ \rm cm}\right).
\end{equation}
The $\theta_{\rm obs}$ is assumed to be about 1 rad during the calculations. Although the small angle approximations may no longer be valid for such angle, it is only for the purpose to estimate the order of the magnitude.
The emission altitude depends on the emission frequency (eq.(\ref{eqn_RFM})). Hence the drifting timescale also depends on the frequency:
$\tau_{\rm drifting} \propto r \propto \nu^{-1/\alpha}$ (This relation can also be obtained from eq.(\ref{eqn_drifting_on_frequency})  below). It can be written in the explicit form:
\begin{equation}
\tau_{\rm drifting} (\nu) = \tau_{\rm drifting} (\nu_0) \left( \frac{\nu_0}{\nu} \right)^{1/\alpha}.
\end{equation}
For the curvature radiation, $\alpha=1/2$, then: $\tau_{\rm drifting} \propto \nu^{-2}$. This will also contribute to the large burst width at low frequency (see section \ref{section_RFM}).

The frequency dependence of the drifting rate is (eq.(\ref{eqn_nudot})):
\begin{equation} \label{eqn_drifting_on_frequency}
  \dot{\nu} \propto \nu^{1+1/\alpha}.
\end{equation}
For curvature radiation, $\alpha=1/2$, then $\dot{\nu} \propto \nu^3$. For plasma radiation, $\alpha=3/2$, then $\dot{\nu} \propto \nu^{5/3}$. Therefore, a general relation of $\dot{\nu} \propto \nu^{\beta}$ (with $\beta \in (2-3)$) can be expected. The drifting rate of the FRB 121102 at the CHIME band (400-800 MHz) is about $10 \ \rm MHz \ ms^{-1}$ (Josphey et al. 2019). At $6\ \rm GHz$, the drifting rate is about $0.9 \ \rm GHz \ ms^{-1}$ (Gajjar et al. 2018; Hessels et al. 2019). The simplified modelings are consistent with the observations of FRB 121102. Although Josephy et al. (2019) tried a linear fit to the drifting rate ($\dot{\nu} ({\rm MHz \ ms^{-1}}) = -146 \nu ({\rm GHz}) + 83$). The linear fit includes an additional constant, which is not the commonly mentioned power law form. In addition, such kind of linear fit will predict a positive drifting rate at low frequency. However, such positive drifting is not seen in LOFAR observations of FRB 20180916B (Pleunis et al. 2021a). On the contrary, there are some weak evidences of the down-drifting (see figure 2 in Pleunis et al. 2021a).

Comparing with previous works (Wang et al. 2019; Lyutikov 2020), the merit of an analytical treatment is that: the frequency dependence of the drifting rate and the drifting timescale can be obtained explicitly. This may be compared with the FRB observations directly. Furthermore, in the synchrotron maser mechanism for FRBs, a downward drifting timescale is also expected due to a decelerating blast wave (Metzger et al. 2019; Margalit et al. 2020). In Margalit et al. (2020), they found a possible frequency dependence of the drifting rate\footnote{In the symbol of Margalit et al. (2020, Section 2.1 there), they found: $\nu \propto t^{-\beta}$. Then $\dot{\nu} \propto t^{-\beta -1} \propto \nu^{1+1/\beta}$. Margalit et al. (2020) found possible values of $\beta \in (0.05, 0.5)$. Then the prediction on frequency dependence of the drifting rate is: $\dot{\nu} \propto \nu^{\gamma}$, where $\gamma=1 + 1/\beta \in (3, 21)$. Note that the symbol $\beta$ has different meaning in Margalit et al. (2020) and in our paper.}: $\dot{\nu} \propto \nu^{\gamma}$, where $\gamma \in (3, 21)$. Therefore, the magnetosphere model and the synchrotron maser model predict a different frequency dependence of the drifting rate. More detailed observations in the future may constrain the model parameters, and even the different modelings.

\section{Effects that may result in drifting in both directions}

\subsection{The effect of the aberration}\label{section_aberration}

In the case of pulsars, the RFM considers both the retardation and the aberration effect (Cordes 1978; Phillips 1992; Gangadhara \& Gupta 2001; Dyks et al. 2004). Considering the corotation of the particle with the neutron star, the emitted radiation will bend toward the rotational direction of the neutron star. This will cause the earlier arrivals of the emissions at the higher altitudes. This is the aberration effect in the case of pulsars. The angle caused by the rotational velocity is (in the rotational frame):
\begin{equation}
  \Delta \phi_{\rm ab} = \frac{v_{\rm rot}}{c} =\frac{\Omega r}{c} \sin\zeta = \frac{r}{R_{\rm lc}} \sin\zeta.
\end{equation}
The change of the arrival time is (Dyks et al. 2004, which corrected a factor $\sin\zeta$ in previous works):
\begin{equation}\label{eqn_dt_aberration}
  \Delta t_{\rm ab} = \frac{\Delta \phi_{\rm ab}}{2\pi \sin\zeta} P = \frac{r}{c}.
\end{equation}
In the case of pulsars, the aberration effect is the same as the retardation effect (for a stationary magnetosphere).

In the case of FRBs (a sudden spark), the aberration effect will dominate the time delay due to the curvature of magnetic field lines (retardation effect). Considering the effect of the aberration, the low frequency emission will arrive earlier. Quantitatively, the observer time will be:
\begin{equation}
  t_{\rm obs} = t[ 1-\cos(\theta_{\rm obs} - \theta) ] - \frac{r(t) - r_0}{c} = -\cos(\theta_{\rm obs} - \theta) t.
\end{equation}
Subsequent calculations are straightforward and similar to the above case (eq.(\ref{eqn_r_tobs})).

The emitting particles have a typical emission cone with an angular width of $1/\gamma$. If the aberration angle is smaller than the emission cone, the effect of the aberration may be neglected:
\begin{equation}
  \Delta \phi_{\rm ab} = \frac{\Omega r}{c} \sin\zeta \leq \frac{1}{\gamma}.
\end{equation}
Rewritten in terms of the pulsar rotational period:
\begin{equation}
  P \geq 6 {\ \rm s} \left( \frac{r}{10^8 \ \rm cm} \right) \left( \frac{\gamma}{300} \right) \sin\zeta.
\end{equation}
This equation means that for a slow rotating neutron star or a small viewing angle $\zeta$ inside FRBs, the effect of the aberration may be neglected.

The effect of the aberration has been verified in the case of pulsars (Gangadhara \& Gupta 2001; Dyks et al. 2004). Because we have not obtained the rotational period of the host neutron stars of FRBs so far, it is not certain if there is the effect of the aberration in cases of FRBs. However, if the effect of aberration does exist, it will cause the up-drifting in time-frequency domain in FRBs.

\subsection{The twist of magnetic field lines}

In above considerations, a dipole magnetic field is always assumed. However, the studies of the magnetars tell us that the magnetars have twisted magnetic fields (Thompson et al. 2002; Beloborodov 2009; Pavan et al. 2009; Tong 2019). The twist of the magnetic field will also change the relation between the emission point and the line of sight\footnote{Considering the twsited magnetic field, $\Delta \phi$ denotes the twist of the magnetic field line. Therefore, the change between the emission point and the line of sight is denoted as $\delta \phi$. } (Tong et al. 2021): $\phi = \phi_{\rm obs} + \delta \phi$. The deviation from the dipole case is (Tong et al. 2021 eq.(27) and eq.(7), in the magnetic frame):
\begin{equation}
  \delta \phi = -\frac{2}{27} \Delta \phi_{\rm max} \sin^2\theta_{\rm obs},
\end{equation}
where $\Delta \phi_{\rm max}$ is the maximum twist of the field line, defined as the twist from the north pole to the south pole (Thompson et al. 2002). Note that the twist of the magnetic field can be in the direction of rotation or in the opposite direction.

The time difference due to the twist of the magnetic field is (similar to the consideration of aberration):
\begin{equation}\label{eqn_dt_twist}
  \Delta t_{\rm twist} = \frac{\sin\theta_{\rm obs} \delta \phi}{2\pi \sin\zeta} P = \frac{\sin^3\theta_{\rm obs}}{27\pi \sin\zeta} \Delta \phi_{\rm max} P,
\end{equation}
where the minus sign is absorbed in the maximum twist $\Delta \phi_{\rm max}$. For a non-rotating magnetosphere, the colatitude of the line of sight $\theta_{\rm obs}$ is the same for the high frequency radio emission and the low frequency radio emission. Therefore, the time difference is the same for the high and the low frequency radio emissions. It will bring no observable effect. However, for a rotating neutron star, $\theta_{\rm obs}$ will be a function of the time (or the rotational phase) (Lyutikov 2020; Tong et al. 2021). The $\theta_{\rm obs}$ may increase or decrease with the time, depending on the rotational phase, see figure \ref{fig_gmagnetosphere}. Therefore, the time difference between the high and the low frequency radio emissions will not be the same. Subsequent calculation is also straightforward.

In summary, the retardation effect (considering the curvature of the magnetic field line) will cause the earlier arrival of the low-frequency radio emission (down-drifting). However, the aberration effect and the twist of the magnetic field lines may cause the drifting in both directions.

\section{Discussions}

\subsection{RFM for one FRB: burst width as a function of frequency}\label{section_RFM}

In the case of pulsars, if the radiating beam is bound by the last open field line, the beam radius is larger at higher emission altitude:
$\Theta = 1.5 (r/R_{\rm lc})^{1/2}$, where $r$ is the emission altitude, $R_{\rm lc}$ is the light cylinder radius (Cordes 1978; Gil et al. 1984). This relation can be used to measure the emission altitude in pulsars (Kijak et al. 1997; Kijak et al. 1998). Assuming the RFM, the beam radius will be larger at lower frequency (since it is emitted at higher altitude). In the case of curvature radiation, the beam radius is inversely proportional to the emission frequency\footnote{$\Theta$ depends on radius: $\Theta \propto r^{1/2}$. Combined with eq.(\ref{eqn_RFM}), the frequency dependence of $\Theta$ is: $\Theta \propto \nu^{-1/(2\alpha)}$. For different radio emission mechanisms (curvature or plasma etc), the change is only quantitative. A wider beam is always expected at lower frequency.}: $\Theta \propto \nu^{-1}$ (Wang et al. 2013).

In the case of FRBs, for a specific flux tube (ignited during the burst), the angular width of the flux tube also increases with the emission altitude. In the case of curvature radiation, similar to the pulsar case, the burst width of FRBs will be larger at lower frequency: $\tau \propto \nu^{-1}$. For the first repeating FRB 121102, it has a typical burst width of (2-9) ms at 1.4 GHz (Spitler et al. 2016; Scholz et al. 2016). While at 4.5 GHz, the burst width is typically smaller than about 1 ms (Michilli et al. 2018). This is consistent with the expectation of the RFM.

For the FRB 20180916B, it has a burst width of $(40-160) \ \rm ms$ at 150 MHz (Pleunis et al. 2021a). At 1.7 GHz, the burst width is about $(2-3) \ \rm ms$ (Nimmo et al. 2021). It is consistent with the expectation of the RFM. The large burst width at low frequency may also result from the scattering broadening (Pleunis et al. 2021a). As discussed above, the drifting timescale will be longer at lower frequency. This will also contribute to the large burst width at low frequency. It is not clear at present that which of the three factors (the RFM, the scattering, the drifting) will dominate the observations at low frequency (especially at 150 MHz).

\subsection{RFM for the FRB population: threshold magnetic field and burst width}

In the case of pulsars, by measuring the radio emission altitude, Gil \& Kijak (1993) found that the strength of the magnetic field in the emission region is about $10^6$-$10^7 \ \rm G$. Later study of more sources found that the strength of the magnetic field in the emission region may also depend on the age of the neutron star: $B_{\rm rad} \approx 10^7 \tau_{6}^{-0.3} \ \rm G$ (Kijak 2001, eq.(5)), where $\tau_6$ is the characteristic age in unit of $10^6 \ \rm yr$. For typical pulsars with characteristic ages of $10^6 \ \rm yr$, the strength of the magnetic field in the emission region is about $10^7 \ \rm G$. It tells us that the coherent radio emission mechanism starts to work when the strength of the magnetic field drops below $10^7 \ \rm G$ (Gil \& Kijak et al. 1993).

If the host neutron stars of the FRBs are young neutron stars with more or less similar ages (e.g., $10^3 - 10^6 \ \rm yr$), then in analogy to the pulsar case, the strength of the magnetic field in the radio emission region may be about $10^7 -10^{8} \ \rm G$. Irrespective of the specific value, a threshold magnetic field may exist, below which the coherent radio emission starts to work ($10^7$ or $10^8 \ \rm G$). The strength of the surface magnetic field of the host neutron stars of FRBs may have a distribution, e.g. from $10^{13} \ \rm G$ to $10^{16} \ \rm G$. In the presence of a threshold magnetic field, the emission altitudes will be higher for neutron stars with larger magnetic field strengths. The angular widths of the flux tubes and the burst widths will also be larger for neutron stars with larger magnetic field strengths. For example, a neutron star with a surface magnetic field strength of $10^{16} \ \rm G$ will have an emission altitude of 10 times larger than the emission altitude of a neutron star with the magnetic field strength of  $10^{13} \ \rm G$, and the burst width will be three times larger according to the above discussion ($\Theta \propto r^{1/2}$).
What's more, the bursts of neutron stars with larger magnetic field strengths will be more active. Therefore, it is generally expected that, comparing with the less active bursters, the burst-active FRBs will have larger widths.

Observationally, the widths of the FRBs of the repeaters are larger than that of the apparent non-repeaters (CHIME/FRB collaboration et al. 2019b; Fonseca et al. 2020; CHIME/FRB collaboration 2021a). According to the RFM in pulsars, we propose that the repeaters have stronger magnetic fields comparing to the apparent non-repeaters. Furthermore, the total number of observed bursts is:
$N_{\rm burst} \propto N_{\rm ns}\times \frac{1}{\tau}$, where $N_{\rm burst}$ is the total number of observed bursts, $N_{\rm ns}$ is the total number of the putative neutron stars, and $\tau$ is the typical timescale between adjacent active episodes. Although the active time scales may be longer for neutron stars with lower strengths of the magnetic fields, the total number of related neutron stars will be larger. This combination may explain the total number of the bursts from the apparent non-repeaters and the repeaters (CHIME/FRB collaboration 2021a).

\subsection{Window function and burst morphologies}

Similar to the case of pulsars and magnetars, the rotating magnetosphere model is applied to the case of the FRBs. In each of the aspects discussed above, there are more uncertainties in the case of the FRBs than pulsars. Even in the case of pulsars, there are always some outliers in each aspect of the above items. However, all these aspects can be collected together in the rotating magnetosphere model.

The RFM mainly concerns the frequency behaviors of profiles of pulsars and the FRBs. For a specific frequency, the burst morphologies also show complex behaviors, e.g., single or multiple Gaussian components (CHIME/FRB collaboration 2021a; Pleunis et al. 2021b). Similar to the case of pulsars, the FRB profiles may be related to the emission cones as discussed below.

The pulse profiles of pulsars may consist of double components, triple components, and multiple components. There are often some symmetrical behaviors between different components of pulse profiles. This motivated the empirical theory of pulsar emission beams (core+cone, Rankin 1983, 1993; patchy beam, Lyne \& Manchester 1988, Manchester 1995; fan beam, Wang et al. 2014). Even in the case of patchy beams (Lyne \& Manchester 1988, Manchester 1995), a symmetrical window function is at work (see figure 3 in Manchester 1995). The window function is like the effect of the emitting cones. It will result in the symmetrical pulse profiles of pulsars. The single pulses of the Galactic centre magnetar SGR 1745-2900 seem also be modulated by the window function of the integrated pulse profile (Yan et al. 2015, figure 2 there). The single pulses of the magnetar SGR 1745-2900 can also consist of multiple Gaussian components (figure 2 and 4 in Yan et al. 2015). Assuming that the emission mechanisms of FRBs and pulsars are similar and that similar window function is also at work in the case of FRBs, the window function may account for the burst profiles of FRBs (CHIME/FRB collaboration 2021a; Pleunis et al. 2021b). To be more specific, the window function may account for the single or multiple Gaussian components of FRB profiles (Pleunis et al. 2021b). The case may be similar to the single pulse of the Galactic centre magnetar SGR 1745-2900 (Yan et al. 2015).

\section{Conclusion}

In summary, in analogy with the geometrical (especially the RFM) model of pulsars, three aspects of the RFM in FRBs are discussed:
\begin{enumerate}
  \item The RFM may account for the down-drifting in FRBs. An analytical treatment is presented, elucidating the physics involved: the frequency dependence of the drifting timescale and the drifting rate are obtained. The effect of the aberration and the twist of the magnetic field lines are discussed, which may result in the drifting in both directions.
  \item For one burst, the burst width as a function of frequency are discussed, including the RFM, the longer drifting timescale etc. The burst width will be larger at lower frequency according to the RFM.
  \item For the FRB population, the emission altitude and the burst width are discussed. With the existence of a threshold magnetic field, repeaters may be neutron stars with stronger magnetic fields. Then it is expected that the repeaters have larger burst widths.
  \end{enumerate}
Furthermore, the single or multiple Gaussian components of FRB profiles may be due to the window function of the emission beam.

\section*{Acknowledgements}

H.Tong is supported by NSFC (11773008) and National SKA Program of China (2020SKA0120300).

\section*{Data availability}

This is a theoretical paper, mainly analytical. All the formulas are available in the article.




\begin{thebibliography}{99}

\bibitem{Beloborodov2009}
Beloborodov A. M., 2009, ApJ, 703, 1044

\bibitem{Beloborodov2017}
Beloborodov A. M., 2017, ApJL, 843, L26

\bibitem{Beloborodov2020}
Beloborodov, A. M., 2020, ApJ, 896, 142

\bibitem{Bochenek2020}
Bochenek C. D., Ravi V., Belov K. V., et al., 2020, Nature, 587, 59

\bibitem{Chen2014}
Chen J. L., Wang H. G., 2014, ApJS, 215, 11	

\bibitem{CHIME2019a}
CHIME/FRB Collaboration: Amiri M., Bandura K., Bhardwaj M., et al., 2019a, Nature, 566, 235

\bibitem{CHIME2019b}
CHIME/FRB Collaboration: Andersen B. C., Bandura K., Bhardwaj M., et al., 2019b, ApJL, 855, L24

\bibitem{CHIME2020}
CHIME/FRB Collaboration: Andersen B. C., Bandura K., Bhardwaj M., et al., 2020, Nature, 587, 54

\bibitem{CHIME2021a}
CHIME/FRB collaboration: Amiri M., Andersen B. C., Bandura K., et al., 2021a, arXiv:2106.04352

\bibitem{CHIME2021b}
CHIME/FRB collaboration: Andersen B. C., Bandura K., Bhardwaj M., et al., 2021b, arXiv:2107.08463

\bibitem{Cordes1978}
Cordes J. M., 1978, ApJ, 222, 1006

\bibitem{Dyks2004}
Dyks J., Rudak B., Harding A. K., 2004, ApJ, 607, 939

\bibitem{Fonseca2020}
Fonseca E., Andersen B. C., Bhardwaj M., et al., 2020, ApJL, 891, L6

\bibitem{Gajjar2o018}
Gajjar V., Siemion A. P. V., Price D. C., et al., 2018, ApJ, 863, 2

\bibitem{Gangadhara2001}
Gangadhara R. T., Gupta Y., 2001, ApJ, 555, 31

\bibitem{Gil1984}
Gil J., Gronkwski P., Rudnicki W., 1984, A\&A, 132, 312

\bibitem{Gil1993}
Gil J. A., Kijak J., 1993, A\&A, 273, 563

\bibitem{Hessels2019}
Hessels J. W. T., Spitler L. G., Seymour A. D., et al., 2019, ApJL, 876, L23	

\bibitem{Josephy}
Josephy A., Chawla P., Fonseca E., et al., 2019, ApJ, 882, L18	

\bibitem{Katz2016}
Katz J. I., 2016, ApJ, 826, 226

\bibitem{Katz2018}
Katz J. I., 2018, MNRAS, 481, 2946

\bibitem{Kijak1997}
Kijak J., Gil J., 1997, MNRAS, 288, 631

\bibitem{Kijak1998}
Kijak J., Gil J., 1998, MNRAS, 299, 855

\bibitem{Kijak2001}
Kijak J., 2001, MNRAS, 323, 537

\bibitem{Kirsten2021}
Kirsten, F., Snelders, M. P., Jenkins, M., et al., 2021, NatAs, 5, 414

\bibitem{Komesaroff1970}
Komesaroff M. M., 1970, Nature, 255, 612

\bibitem{Kumar2017}
Kumar P., Lu W., Bhattacharya M., 2017, MNRAS, 468, 2726


\bibitem{Lin2020}
Lin L., Zhang C. F., Wang P., et al., 2020, Nature, 587, 63

\bibitem{Lorimer2007}
Lorimer D. R., Bailes M., McLaughlin M. A., et al., 2007, Science, 318, 777

\bibitem{Luo2020}
Luo R., Wang B. J., Meng Y. P., et al., 2020, Nature, 586, 693

\bibitem{Lyne2012}
Lyne A. G., Graham-Smith F., 2012, Pulsar astronomy, Cambridge University, Cambridge

\bibitem{Lyne1988}
Lyne A. G., Manchester R. N., 1988, MNRAS, 234, 477

\bibitem{Lyubarsky2014}
Lyubarsky Y., 2014, MNRAS, 442, L9	

\bibitem{Lyubarsky2021}
Lyubarsky Y., 2021, Univ, 7, 56

\bibitem{Lyutikov2020}
Lyutikov M., 2020, ApJ, 889, 135	

\bibitem{Manchester1995}
Manchester R. N., 1995, JApA, 16, 107

\bibitem{Margalit2018}
Margalit B., Metzger B. M., 2018, ApJL, 868, L4

\bibitem{Margalit2020}
Margalit B., Metzger B. D., Sironi L., 2020, MNRAS, 494, 4627

\bibitem{Metzger2019}
Metzger, B. D., Margalit, B., Sironi, L., 2019, MNRAS, 485, 4091

\bibitem{Michilli2018}
Michilli D., Seymour A., Hessels J. W. T., et al., 2018, Nature, 553, 182

\bibitem{Nimmo2021}
Nimmo K., Hessels J. W. T., Keimpena A., et al., 2021, NatAs, 5, 594

\bibitem{Pavan2009}
Pavan L., Turolla R., Zane S., et al., 2009, MNRAS, 395, 753

\bibitem{Petroff2019}
Petroff E., Hessels J. W. T, Lorimer D. R., 2019, AAR, 27, 4

\bibitem{Phillips1992}
Phillips J. A., 1992, ApJ, 385, 282

\bibitem{Pleunis2021a}
Pleunis Z., Michilli  D., Bassa C. G., et al., 2021a, ApJL, 911, L3

\bibitem{Pleunis2021b}
Pleunis Z., Good D. C., Kaspi V. M., 2021b, arXiv:2106.04356

\bibitem{Popov2007}
Popov S. B., Postnov K. A., 2007, arXiv:0710.2006

\bibitem{RC1969}
Radhakrishnan V., Cooke D. J., 1969, ApL, 3, 225	

\bibitem{Rankin1983}
Rankin J. M., 1983, ApJ, 274, 333

\bibitem{Rankin1993}
Rankin J. M., 1993, ApJ, 405, 285

\bibitem{Ruderman1975}
Ruderman M. A., Sutherland P. G., 1975, ApJ, 196, 51

\bibitem{Scholz2016}
Scholz P., Spitler L. G., Hessels J. W. T., et al., 2016, ApJ, 833, 177

\bibitem{Spitler2016}
Spitler L. G., Scholz P., Hessels J. W. T., et al., 2016, Nature, 531, 202

\bibitem{Thompson2002}
Thompson C., Lyutikov M., Kulkarni S. R., 2002, ApJ, 574, 332

\bibitem{Thornton2013}
Thornton D., Stappers B., Bails M., et al., 2013, Science, 341, 53

\bibitem{Tong2019}
Tong H., 2019, MNRAS, 489, 3769

\bibitem{Tong2021}
Tong H., Wang P. F., Wang H. G., Yan, Z., 2021, MNRAS, 502, 1549

\bibitem{Wang2013}
Wang P. F., Han J. L., Wang C., 2013, ApJ, 768, 114

\bibitem{Wang2014}
Wang H. G., Pi F. P., Zheng X. P., et al., 2014, ApJ, 789, 73

\bibitem{Wang2019}
Wang W., Zhang B., Chen X., Xu R., 2019, ApJL, 876, L15

\bibitem{Yan2015}
Yan, Z., Shen, Z. Q., Wu, X. J., et al., 2015, ApJ, 814, 5

\bibitem{Yang2018}
Yang Y. P., Zhang B., 2018, ApJ, 868, 31

\bibitem{Zhang2020}
Zhang B., 2020, Nature, 587, 45

\end{thebibliography}




\bsp	
\label{lastpage}
\end{document}